\documentclass[conference]{IEEEtran}
\usepackage[latin9]{inputenc}
\usepackage{amsmath}
\usepackage{graphicx}

\makeatletter

\def\ps@IEEEtitlepagestyle{%
  \def\@oddfoot{\mycopyrightnotice}%
}

\def\mycopyrightnotice{%
}
\makeatother

\makeatother

\usepackage{cite}

\ifCLASSINFOpdf
\else
\fi
\usepackage{amsmath}

\usepackage{stfloats}
\begin{document}
%
\title{Minimizing Embedding Distortion with Weighted Bigraph Matching in Reversible Data Hiding}



%
\author{Hanzhou Wu\\
Homepage: hzwu.github.io, Email: wuhanzhou\_2007@126.com\\
This manuscript is a short version for only ensuring timely dissemination of the research.\\
I did not show all the details in this version. However, you may understand the core idea of this work.
}


\maketitle

\begin{abstract}
For a required payload, the existing reversible data hiding (RDH) methods always expect to reduce the embedding distortion as much as possible, such as by utilizing a well-designed predictor, taking into account the carrier-content characteristics, and/or improving modification efficiency etc. However, due to the diversity of natural images, it is actually very hard to accurately model the statistical characteristics of natural images, which has limited the practical use of traditional RDH methods that rely heavily on the content characteristics. Based on this perspective, instead of directly exploiting the content characteristics, in this paper, we model the embedding operation on a weighted bipartite graph to reduce the introduced distortion due to data embedding, which is proved to be equivalent to a graph problem called as \emph{minimum weight maximum matching (MWMM)}. By solving the MWMM problem, we can find the optimal histogram shifting strategy under the given condition. Since the proposed method is essentially a general embedding model for the RDH, it can be utilized for designing an RDH scheme. In our experiments, we incorporate the proposed method into some related works, and, our experimental results have shown that the proposed method can significantly improve the payload-distortion performance, indicating that the proposed method could be desirable and promising for practical use and the design of RDH schemes.
\end{abstract}


%
\IEEEpeerreviewmaketitle

\section{Motivation}
Unlike steganography \cite{fridrich:book}, reversible data hiding (RDH) \cite{ni:HS} allows both the hidden information and the host content to be perfectly reconstructed for a receiver, which is applicable to sensitive scenarios that require no degradation of the host data, such as military and remote sensing.
Both steganography and RDH expect to minimize the embedding distortion when subjected to a fixed payload. Since there has no need to recover the original content, for steganography, one could use such as syndrome-trellis codes (STCs) \cite{fridrich:STCs} and Gibbs construction \cite{fridrich:Gibbs} to minimize or simulate the embedding impact. However, due to the requirement of reversibility, these optimization methods suited to steganography could not be directly applied to RDH, which has motivated us to study the distortion optimization of the RDH in this paper.

As an efficient embedding strategy, histogram shifting (HS) \cite{ni:HS} has been widely utilized in the reported RDH works \cite{tsai:pchs, li:GeneralHS}. For most of the HS-based RDH methods, the data hider should process the cover pixels by using a well-designed pixel prediction and selection rule, so that the generated difference (or prediction-error) histogram is sharply distributed \cite{hzwu:DCSPF, hzwu:ppe}, which can benefit for data embedding. 
Since the pixel prediction and selection procedure often relies heavily on the carrier-content characteristics, the payload-distortion behavior will vary due to the diversity of natural images, which, to a certain extent, has limited the practical use of these RDH methods. 
\begin{figure}[!t]
\centering
\includegraphics[width=3in]{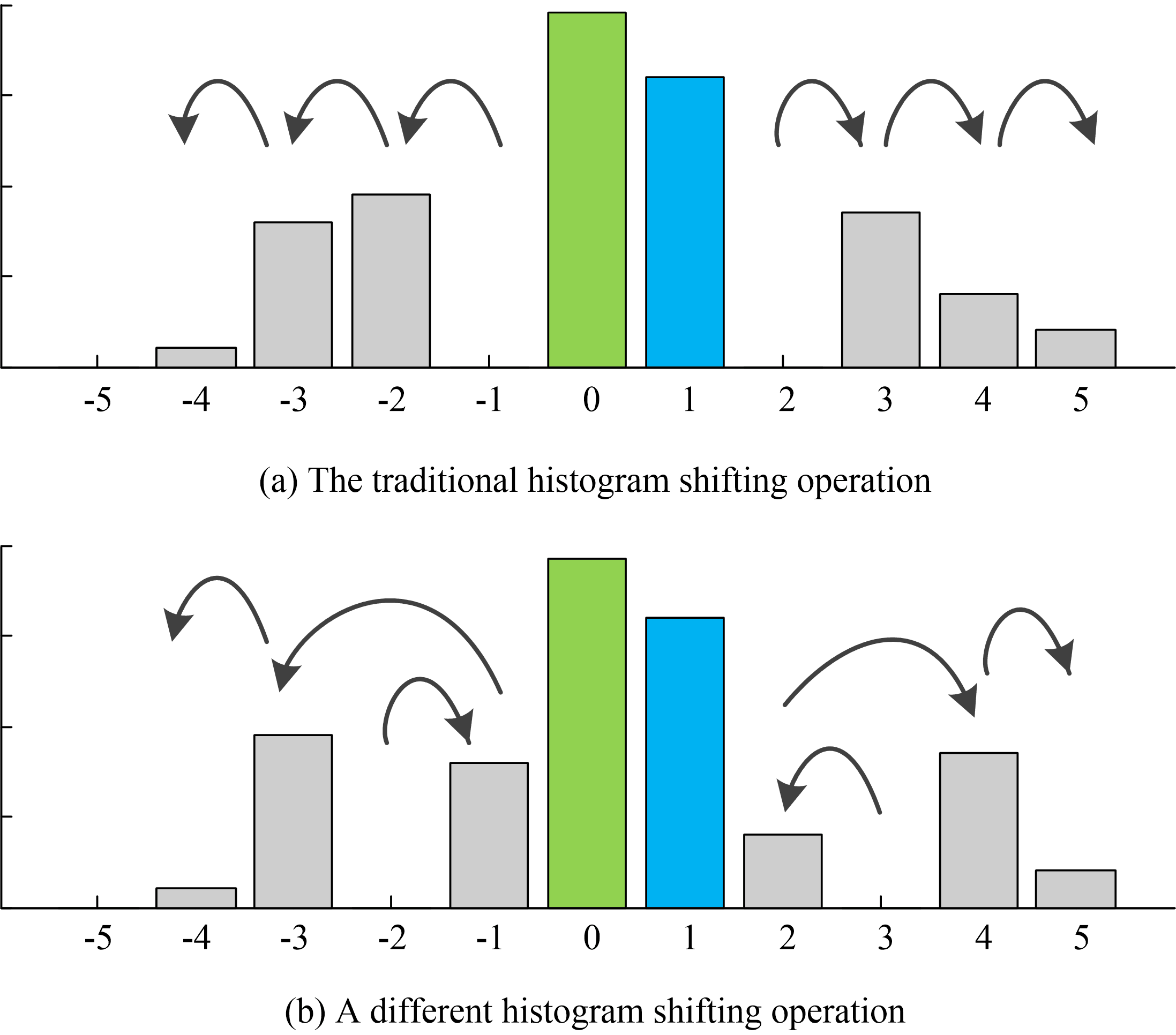}
\caption{Two examples of histogram shifting used in RDH.}
\end{figure}

On the other hand, with a generated histogram, one should choose suitable peak bins (i.e., the bins usually with maximum occurrences) to embed the secret data. Moreover, some of the other histogram bins should be shifted to ensure reversibility. It is quite desirable to choose such peak bins that they can carry the secret data while keep the distortion low. Though the bins shifted for reversibility do not carry the secret data, they often introduce larger distortion than the peak bins. For a \emph{single-layer} embedding, one may easily find the optimal shifting strategy, since a pixel is increased or decreased by at most one. However, when to adopt \emph{multi-layer} embedding, since the existing works shift the bins along the corresponding direction with a fixed step value, the cover pixels may change greatly, resulting in significant degradation of the image quality. 

Based on the above-mentioned perspective, instead of designing a detailed HS-based RDH (that relies heavily on image content), we hope to model the HS-based embedding operation as a general framework. Specifically, we are to optimize the shifting operation when to use multi-layer embedding, so that the distortion can be significantly reduced. In our work, we model the shifting operation on a weighted bipartite graph, in which the vertices represent the histogram bins to be modified and edges indicate the shifting-relationship among the vertices. All edges are assigned with a weight (or cost) to specify the corresponding shifting-distortion. By solving a standard graph matching problem called \emph{minimum weight maximum matching (MWMM)}, we can finally find the optimal histogram shifting strategy, which ensures the minimum distortion.

The rest of this paper are organized as follows. The problem of shifting operation is formulated in Section II. In Section III, we introduce the proposed optimization model of minimizing embedding distortion with weighted bigraph matching. Some experimental results and analysis are provided in Section IV. Finally, we conclude this paper in Section V.
\section{Problem Formulation}
Before we formulate the optimization problem, let us start with an example. As shown in Fig. 1, we use two peak bins ``0'' and ``1'' to hide a message. The traditional RDH method, i.e., Fig. 1 (a), first shifts the other bins along the corresponding direction by a step value of 1. Then, the embedding space reserved by ``-1'' and ``2'' can be exploited to carry the message by shifting ``0'' and ``1'', respectively. This fixed empirical shifting-pattern may be not optimal when a multi-layer embedding is adopted since the prediction-error (PE) of a pixel will become larger. And, there has no work to explicitly demonstrate that this operation will still introduce the lowest distortion. For example, Fig. 1 (b) may be the optimal shifting operation (or that outperforms the traditional one) for a higher-layer embedding. In this paper, we will optimize this shifting operation to reduce the distortion for a multi-layer embedding.

We call $\textbf{x}^{(t)} (t\geq 0)$ the cover image after embedded with $t$ times. Obviously, $\textbf{x}^{(0)}$ is the original image without any hidden bits. For simplicity, let $\textbf{x}^{(t)} = ({x_1}^{(t)}, {x_2}^{(t)}, ..., {x_n}^{(t)}) \in \mathcal{X} = \{\mathcal{I}\}^n$ be an $n$-pixel cover image with the pixel range $\mathcal{I}$, e.g., $\mathcal{I}=\{0,1,...,255\}$ for 8-bit grayscale images. 
For a payload, we will use $\textbf{x}^{(0)}$ and $\textbf{x}^{(t)}$ to generate the marked image $\textbf{x}^{(t+1)} (t\geq 0)$ with HS operation. Our goal is to minimize the distortion $D$ between $\textbf{x}^{(0)}$ and $\textbf{x}^{(t+1)}$.  We here limit ourselves to an additive $D$ as:
\begin{equation}
D(\textbf{x}^{(0)}, \textbf{x}^{(t+1)}) = \sum_{i=1}^{n}\rho_i(\textbf{x}^{(0)}, {x_i}^{(t+1)}),
\end{equation}
where $\rho_i:\mathcal{X}\times \mathcal{I}_i\mapsto R$ exposes the cost of changing ${x_i}^{(0)}$ to ${x_i}^{(t+1)}$. In RDH, we often use the squared error to evaluate the distortion, which can be generalized by Eq. (1). So, in default, we will use mean squared error (MSE) as the measure, i.e.,
\begin{equation}
D(\textbf{x}^{(0)}, \textbf{x}^{(t+1)}) = \frac{1}{n}\cdot\sum_{i=1}^{n}({x_i}^{(0)} - {x_i}^{(t+1)})^2.
\end{equation}

In RDH, we need to predict the pixels to be embedded in $\textbf{x}^{(t)}$, and generate the corresponding pixel prediction-error histogram (PEH). Without the loss of generality, let $h(v)$ be the occurrence of the PEH bin with a value of $v$. Here, we have $-|\mathcal{I}| < v< |\mathcal{I}|$, where $|*|$ represents the size of a set. To hide a message, with the generated PEH, one should shift some PEs to vacate empty bin-positions, and then embed the secret bits by shifting the peak bins into the empty bin-positions. 
Mathematically, let $\mathcal{A}$ and $\mathcal{B}$ denote a set including all PEH bins and 
that contains all \emph{non-zero} occurrence bins, respectively. It means that
$\mathcal{A} = \{v~| -|\mathcal{I}|< v< |\mathcal{I}| \}$ and  $\mathcal{B} = \{v~|~h(v) > 0\} \subset \mathcal{A}$.
For a peak-bin set $\mathcal{P} = \{p_1, p_2, ..., p_m\}\subset \mathcal{B}$, we first find such two injective functions $g_0$ and $g_1$ that, 
$\mathcal{G}_0 = \{g_0(p_1), g_0(p_2), ..., g_0(p_m)\}\subset \mathcal{A}$ and $\mathcal{G}_1 = \{g_1(p_1), g_1(p_2), ..., g_1(p_m)\}\subset \mathcal{A}$, where
$\mathcal{G}_0\cap\mathcal{G}_1=\emptyset$. Then, another injective function $f: \mathcal{B}\setminus\mathcal{P} \mapsto \mathcal{A}\setminus(\mathcal{G}_0\cup\mathcal{G}_1)$ is also required. Suppose the bit-size of message is no more than $\sum_{i=1}^{m} h(p_i)$, for data embedding, according to $f$, we first shift all PEH bins in $\mathcal{B}\setminus\mathcal{P}$ into some bin-positions of $\mathcal{A}\setminus(\mathcal{G}_0\cup\mathcal{G}_1)$. Thereafter, since the bin-positions of $(\mathcal{G}_0\cup\mathcal{G}_1)\setminus\mathcal{P}$ are empty (i.e., with zero occurrence), one can easily embed the secret bits by shifting the bins in $\mathcal{P}$ into the bin-positions of $(\mathcal{G}_0\cup\mathcal{G}_1)$.
\begin{figure}[!t]
\centering
\includegraphics[width=3in]{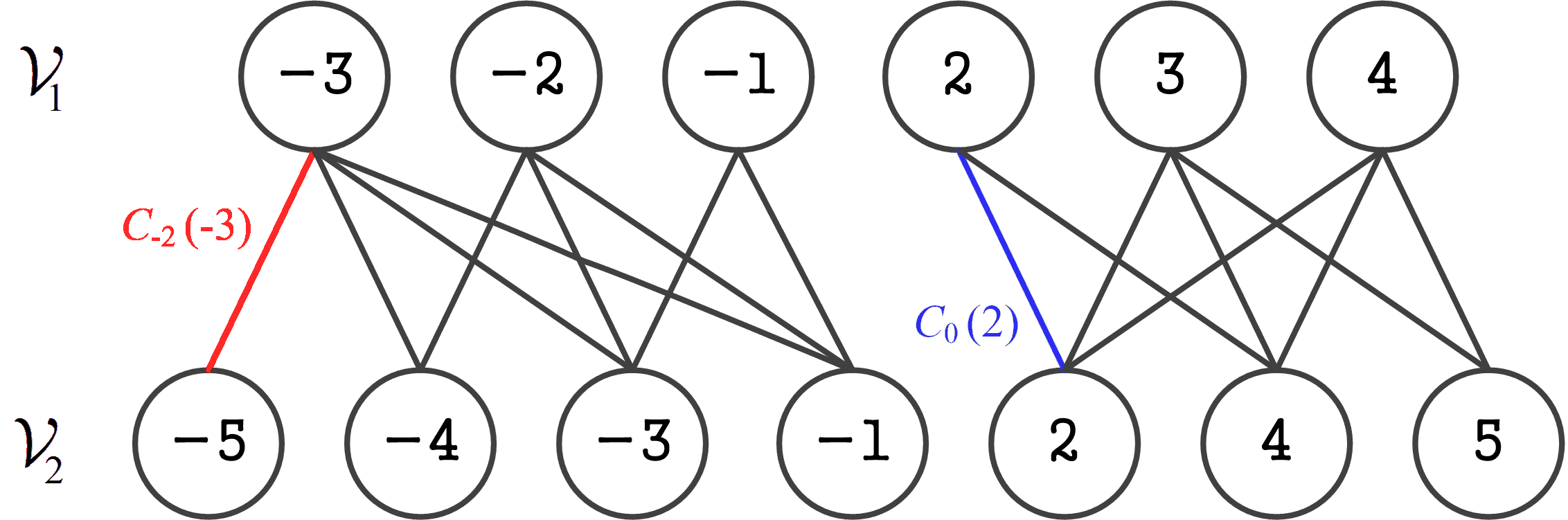}
\caption{An example of weighted bipartite graph.}
\end{figure}
We here take Fig. 1 for explanation. It can be inferred that, $\mathcal{A} = \{-5, -4, ..., 5\}$ and $\mathcal{B} = \{-3, -2, ..., 4\}$.
In both cases, we have $\mathcal{P} = \mathcal{G}_0 = \{0, 1\}$. However, in Fig. 1 (a), we have $\mathcal{G}_1 = \{-1, 2\}$, while in Fig. 1 (b), $\mathcal{G}_1 = \{-2, 3\}$. In Fig. 1 (a), the injective function $f$ maps \{-3, -2, -1, 2, 3, 4\} to \{-4, -3, -2, 3, 4, 5\}, respectively; and in Fig. 1 (b), it maps \{-3, -2, -1, 2, 3, 4\} to \{-4, -1, -3, 4, 2, 5\}, respectively.

Obviously, when $\mathcal{P}$, $g_0$ and $g_1$ are fixed, it is quite desirable to find the best $f$ such that $D$ can be minimized. Once this optimization problem is solved, one can enumerate $\mathcal{P}$, $g_0$ and $g_1$ to minimize the global distortion for a payload since $|\mathcal{P}|$ is often small, e.g., $|\mathcal{P}| = 2$. We will study along this direction.

Let $\textbf{c}^{(t)} = ({c_1}^{(t)}, {c_2}^{(t)}, ..., {c_{n_t}}^{(t)}),n_t\leq n$, represent all the cover pixels to be embedded.
For compactness, we sometimes consider $\textbf{c}^{(t)}$ and ${c_i}^{(t)}$ as the pixel set containing all the cover pixels to be embedded and the $i$-th pixel with a value of ${c_i}^{(t)}$, respectively. Similarly, we denote the prediction of $\textbf{c}^{(t)}$ and its marked version by $\textbf{z}^{(t)} = ({z_1}^{(t)}, {z_2}^{(t)}, ..., {z_{n_t}}^{(t)})$ and $\textbf{s}^{(t)} = ({s_1}^{(t)}, {s_2}^{(t)}, ..., {s_{n_t}}^{(t)})$. Thus, we can find the PEs $\textbf{e}^{(t)} = ({e_1}^{(t)}, {e_2}^{(t)}, ..., {e_{n_t}}^{(t)})$ between $\textbf{c}^{(t)}$ and $\textbf{z}^{(t)}$ by
\begin{equation}
{e_i}^{(t)} = {c_i}^{(t)} - {z_i}^{(t)}, (1\leq i\leq n_t). 
\end{equation}

The relationship of $\textbf{c}^{(t)}$ and $\textbf{s}^{(t)}$ can be described as:
\begin{equation}
{{s_i}^{(t)}}=\begin{cases}
{z_i}^{(t)}+g_{b_k}({e_i}^{(t)}), & \text{ if } {e_i}^{(t)}\in \mathcal{P};\\ 
{z_i}^{(t)}+f({e_i}^{(t)}), & \text{ if } {e_i}^{(t)}\in \mathcal{B}\setminus\mathcal{P}; \\ 
{z_i}^{(t)}+{e_i}^{(t)}, & ~\text{otherwise}. 
\end{cases}
\end{equation}
Here, $b_k=\{0,1\}$ is the $k$-th (current) bit to be embedded.

We use $\textbf{o}^{(t)} = ({o_1}^{(t)}, {o_2}^{(t)}, ..., {o_{n_t}}^{(t)})$ to denote the original pixel values of $\textbf{c}^{(t)}$ in $\textbf{x}^{(0)}$. It is pointed out that, for the pixels not belonging to $\textbf{c}^{(t)}$, the distortion can be roughly considered as fixed since we will not embed secret data into these pixels (though we may alter some pixels prior to embedding, e.g., to empty some LSBs to store the secret key). Therefore, for $(t+1)$-layer ($t \geq 0$) embedding (i.e., to generate $\textbf{x}^{(t+1)}$), our optimization task is
\begin{equation}
D(\textbf{x}^{(0)}, \textbf{x}^{(t+1)}) = \underset{\mathcal{P},g_0,g_1,f}{\text{min}}~~\frac{1}{n}\cdot\sum_{i=1}^{n_t}({s_i}^{(t)}-{o_i}^{(t)})^2+\frac{E}{n},
\end{equation}
where $E$ is a constant. With Eq. (4), we have
\begin{equation}
\begin{split}
\sum_{i=1}^{n_t}({s_i}^{(t)}-{o_i}^{(t)})^2 &= \sum_{{e_i}^{(t)}\in\mathcal{P}}(g_{b_k}({e_i}^{(t)})+{z_i}^{(t)}-{o_i}^{(t)})^2\\
&+\sum_{{e_i}^{(t)}\in\mathcal{B}\setminus\mathcal{P}}(f({e_i}^{(t)})+{z_i}^{(t)}-{o_i}^{(t)})^2\\
&+\sum_{{e_i}^{(t)}\notin\mathcal{B}}({e_i}^{(t)}+{z_i}^{(t)}-{o_i}^{(t)})^2.
\end{split}
\end{equation}

For RDH, the secret bits can be orderly embedded into $\textbf{c}^{(t)}$ since $\textbf{c}^{(t)}$ can be generated by a key or some specified rule. When $\mathcal{P}$, $g_0$ and $g_1$ are fixed, with the secret message, one can consider $\sum_{{e_i}^{(t)}\in\mathcal{P}}(g_{b_k}({e_i}^{(t)})+{z_i}^{(t)}-{o_i}^{(t)})^2$ as fixed. Actually, even without the message, one can use a random bitstring to simulate it so that, $\sum_{{e_i}^{(t)}\in\mathcal{P}}(g_{b_k}({e_i}^{(t)})+{z_i}^{(t)}-{o_i}^{(t)})^2$ can be roughly estimated. 
It is obvious that,
$\sum_{{e_i}^{(t)}\notin\mathcal{B}}({e_i}^{(t)}+{z_i}^{(t)}-{o_i}^{(t)})^2$ is fixed as well. Therefore, we have
\begin{equation}
\begin{split}
D(\textbf{x}^{(0)}, \textbf{x}^{(t+1)};\mathcal{P},g_0,g_1) & = \\
\underset{f}{\text{min}}~~&\frac{1}{n}\cdot\sum_{i=1}^{n_t}({s_i}^{(t)}-{o_i}^{(t)})^2+\frac{E}{n},
\end{split}
\end{equation}
which actually requires us to minimize
\begin{equation}
L = \sum_{{e_i}^{(t)}\in\mathcal{B}\setminus\mathcal{P}}(f({e_i}^{(t)})+{z_i}^{(t)}-{o_i}^{(t)})^2.
\end{equation}

Now we need to propose an efficient algorithm to find such $f$ that Eq. (8) can be minimized for fixed $\mathcal{P}$, $g_0$ and $g_1$.
Thereafter, by enumerating $\mathcal{P}$, $g_0$ and $g_1$, we can find the optimal $\mathcal{P}$, $g_0$, $g_1$ and $f$ for $\textbf{x}^{(t+1)}$ based on $\textbf{x}^{(0)}$ and $\textbf{x}^{(t)}$.
\section{Minimizing Embedding Distortion with Weighted Bigraph Matching}
In this section, we will introduce the method called \emph{weighted bigraph matching} to minimize $L$ in Eq. (8). 
\subsection{Model Derivation}
Without the loss of generality, we rewrite $f$ as:
\begin{equation}
f(x)=x+\triangle_f(x),
\end{equation}
where $\triangle_f(x)$ has no need to be injective. Therefore, we have
\begin{equation}
L = \sum_{{e_i}^{(t)}\in\mathcal{B}\setminus\mathcal{P}}(\triangle_f({e_i}^{(t)})+{c_i}^{(t)}-{o_i}^{(t)})^2.
\end{equation}

In RDH, to avoid the underflow/overflow problem, we need to adjust the pixels with boundary values into the reliable range in advance. Therefore, $\triangle_f(x)$ should be bounded at the very beginning so that one can process the boundary pixels in advance, i.e., 
\begin{equation*}
-T \leq \triangle_f(x) \leq T,
\end{equation*}
where $T$ is a positive integer threshold, e.g., $T = 1$. Actually, $g_0(x)$ and $g_1(x)$ should be bounded as well. 
For simplicity, it is considered that $|x-g_0(x)|\leq T$ and $|x-g_1(x)|\leq T$.

Let $\{y_1, y_2, ..., y_{|\mathcal{B}\setminus\mathcal{P}|}\}$ denote all the elements in $\mathcal{B}\setminus\mathcal{P}$. Eq. (10) can be therefore rewritten as:
\begin{equation}
L = \sum_{j=1}^{|\mathcal{B}\setminus\mathcal{P}|}J(y_j; f),
\end{equation}
where
\begin{equation*}
\begin{split}
J(y_j; f) &= \sum_{{e_i}^{(t)}=y_j}(\triangle_f(y_j)+{c_i}^{(t)}-{o_i}^{(t)})^2\\
&=\sum_{{e_i}^{(t)}=y_j} \sum_{k=-T}^{T} \delta(\triangle_f(y_j), k)\cdot(k+{c_i}^{(t)}-{o_i}^{(t)})^2\\
&=\sum_{k=-T}^{T} \delta(\triangle_f(y_j), k)\sum_{{e_i}^{(t)}=y_j}(k+{c_i}^{(t)}-{o_i}^{(t)})^2\\
&=\sum_{k=-T}^{T} \delta(f(y_j)-y_j, k)\cdot C_k(y_j),
\end{split}
\end{equation*}
where $\delta(x, y) = 1$ if $x = y$, otherwise $\delta(x, y) = 0$; and,
\begin{equation}
C_k(y_j) = \sum_{{e_i}^{(t)}=y_j}(k+{c_i}^{(t)}-{o_i}^{(t)})^2.
\end{equation}

Now, our problem is to find the best injective function $f$ for Eq. (11), 
which can be addressed by applying the weighted bigraph matching method introduced in the following.
\subsection{Weighted Bigraph Matching}
For RDH, every element in $\mathcal{B}\setminus\mathcal{P}$ should be matched by exactly one element (unique) in 
$\mathcal{A}\setminus(\mathcal{G}_0\cup\mathcal{G}_1)$ according to $f$. Note that, in RDH, we often have $|\mathcal{B}\setminus\mathcal{P}| \leq |\mathcal{A}\setminus(\mathcal{G}_0\cup\mathcal{G}_1)|$. On the other hand, we expect to find such an optimal matching scheme $f_{\text{opt}}$ that Eq. (11) can be minimized. Accordingly, our optimization task is finally generalized as:
\begin{equation}
f_{\text{opt}} = \underset{f}{\text{arg min}} 
\sum_{j=1}^{|\mathcal{B}\setminus\mathcal{P}|}\sum_{k=-T}^{T} \delta(f(y_j)-y_j, k)\cdot C_k(y_j),
\end{equation}
subject to
\begin{equation}
\sum_{k=-T}^{T} \delta(f(y_j)-y_j,k) = 1,~\text{for all}~1\leq j\leq |\mathcal{B}\setminus\mathcal{P}|.
\end{equation}

Obviously, with Eq. (12), all possible $C_k(y_j)$ can be easily determined in advance. Without loss of generality, we will use $\{q_1, q_2, ..., q_{|\mathcal{A}\setminus(\mathcal{G}_0\cup\mathcal{G}_1)|}\}$ to represent the elements in $\mathcal{A}\setminus(\mathcal{G}_0\cup\mathcal{G}_1)$.
We are to model the optimization problem of Eq. (13) on a \emph{weighted bipartite graph}. A \emph{bipartite graph}, or \emph{bigraph}, is a graph whose vertices can be partitioned into such two disjoint sets $\mathcal{V}_1$ and $\mathcal{V}_2$ that all edges connect a vertex in $\mathcal{V}_1$ and one in $\mathcal{V}_2$. If all edges in a bipartite graph are assigned to a weight, it is named as a \emph{weighted bipartite graph} (or \emph{weighted bigraph}).

To build a weighted bipartite graph, we first denote the two disjoint sets by $\mathcal{V}_1 = \mathcal{B}\setminus\mathcal{P}$ and 
$\mathcal{V}_2 = \mathcal{A}\setminus(\mathcal{G}_0\cup\mathcal{G}_1)$. With Eq. (14), for every possible index-pair $(i, j)$, if $|y_j-q_i|\leq T$, we assign an edge between $y_j$ and $q_i$ in the bipartite graph. 
It indicates that, it is possible that $f_{\text{opt}}(y_j) = q_i$. 
Meanwhile, all edges will be assigned with the corresponding weights. Specifically, if there exists an edge between $y_j$ and $q_i$, the assigned weight should be $C_{q_i-y_j}(y_j)$, meaning that, if $f_{\text{opt}}(y_j) = q_i$, the corresponding shifting-distortion should be $C_{q_i-y_j}(y_j)$.
\begin{figure}[!t]
\centering
\includegraphics[width=3in]{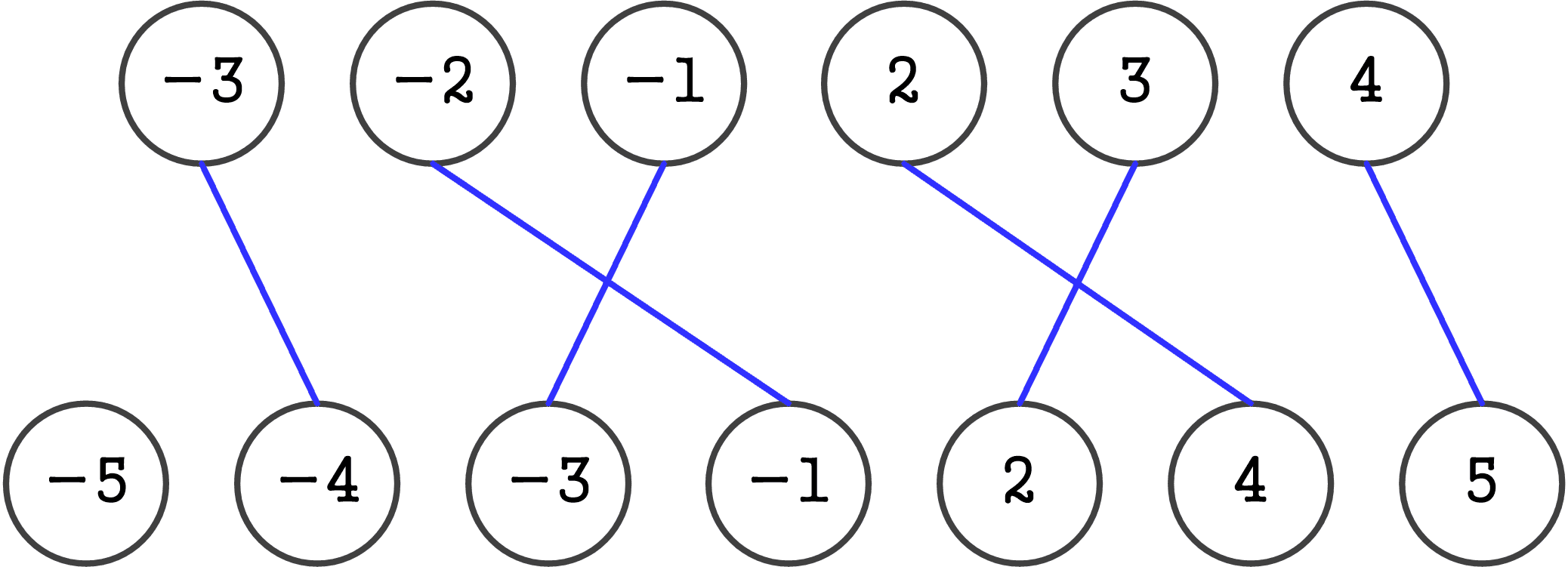}
\caption{An example of maximum matching for Fig. 2.}
\end{figure}

We take Fig. 1 for example. Suppose that $T = 2$, $\mathcal{P} = \mathcal{G}_0 = \{0, 1\}$ and $\mathcal{G}_1 = \{-2, 3\}$, 
we have $\mathcal{V}_1 = \{-3, -2, -1, 2, 3, 4\}$ and $\mathcal{V}_2 = \{-5, -4, -3, -1, 2, 4, 5\}$. The weighted bigraph can be therefore built as Fig. 2. 
In Fig. 2, the weight of each edge can be determined according to Eq. (12), e.g., 
the weight between ``-3'' (in $\mathcal{V}_1$) and ``-5'' (in $\mathcal{V}_2$) is $C_{-2}(-3)$.

A \emph{matching} $\mathcal{M}$ of a bigraph is a set of \emph{non-adjacent} edges, i.e., there has no two edges in $\mathcal{M}$ sharing a common vertex. 
A \emph{maximum matching} $\mathcal{M}$ of a bigraph is such a matching that it is not a subset of any other matching. In other words, a matching $\mathcal{M}$ of a bigraph is maximum if every edge in the bigraph has a \emph{non-empty} intersection with one edge in $\mathcal{M}$. 

As shown in Fig. 3, we show an example of maximum matching for the bigraph built in Fig. 2, e.g., in Fig. 3, ``-2'' is matched by ``-1''. Obviously, in a bigraph, there may be many candidates of maximum matching. A maximum matching guarantees that, there has no two edges sharing the same vertex, and the total number of edges in the matching is maximum. For any maximum matching $\mathcal{M}_{\text{max}}$ of a bigraph, there must be $|\mathcal{M}_{\text{max}}|\leq \text{min}\{|\mathcal{V}_1|, |\mathcal{V}_2|\}$.

As $f_{\text{opt}}$ is injective, we can infer that, in the corresponding weighted bigraph, $f_{\text{opt}}$ corresponds to such a matching $\mathcal{M}_{\text{opt}}$ that $|\mathcal{M}_{\text{opt}}| = |\mathcal{V}_1|$, where $\mathcal{M}_{\text{opt}}$ is also a maximum matching since $|\mathcal{M}_{\text{opt}}| = |\mathcal{V}_1| \geq \text{min}\{|\mathcal{V}_1|, |\mathcal{V}_2|\} \geq |\mathcal{M}_{\text{max}}|$, i.e.,

\textbf{Proposition 1.} $f_{\text{opt}}$ corresponds to such a maximum matching $\mathcal{M}_{\text{opt}}$ that $|\mathcal{M}_{\text{opt}}| = |\mathcal{V}_1|$, where $\mathcal{V}_1 = \mathcal{B}\setminus\mathcal{P}$.

Moreover, according to Eq. (13), $f_{\text{opt}}$ requires that, the sum of edge-weights in $\mathcal{M}_{\text{opt}}$ should be the minimum. Therefore, to find $f_{\text{opt}}$, we have to determine
\begin{equation}
\mathcal{M}_{\text{opt}} = \underset{|\mathcal{M}|=|\mathcal{V}_1|}{\text{arg min}} 
\sum_{(y_j,q_i)\in \mathcal{M}, y_j\in \mathcal{V}_1, q_i\in \mathcal{V}_2} C_{q_i-y_j}(y_j).
\end{equation}
Namely,

\textbf{Proposition 2.} $\mathcal{M}_{\text{opt}}$ has the minimum sum of edge-weights.

In graph theory, for a weighted bipartite graph, a \emph{minimum weight maximum matching (MWMM)} is defined as a maximum matching where the sum of the weights associated to edges in the matching has a minimum value, which can be solved by Hungarian algorithm optimized with a time complexity of $O(V^3)$ \cite{roberts:math}. Therefore, we can determine $\mathcal{M}_{\text{opt}}$ in the weighted bigraph with Hungarian algorithm, and then easily construct $f_{\text{opt}}$ with $\mathcal{M}_{\text{opt}}$. We will not introduce the Hungarian algorithm in detail. We here refer a reader to \cite{roberts:math}.

When to produce $\textbf{x}^{(t+1)}$ (based on $\textbf{x}^{(0)}$ and $\textbf{x}^{(t)}$), the traditional HS operation only corresponds to a maximum matching, it may not ensure the minimum distortion. Therefore, in theory, the payload-distortion of an RDH scheme equipped with our optimization method will not be worse than that with the traditional operation. If the traditional HS strategy is optimal in some cases, our method will find it out.

\subsection{Complexity Analysis}
We have introduced the method to find the best $f$ for fixed $\mathcal{P}$, $g_0$ and $g_1$. To find the global-optimal strategy, we have to further enumerate all possible combinations between $\mathcal{P}$, $g_0$ and $g_1$. Since in applications, $|\mathcal{P}|$ is often small, one can easily find all possible $\mathcal{P}$. For example, if $|\mathcal{P}| = 2$, the time complexity is $O(|\mathcal{B}|^2)$, where $|\mathcal{B}| \ll |\mathcal{A}|$ (e.g., $|\mathcal{B}| = 40$ and $|\mathcal{A}| = 511$) since the generated PEH is often sharply distributed (centered at zero-bin). Actually, as $\sum_{p\in \mathcal{P}} h(p)$ should be no less than the bit-length of required payload, the total number of usable $\mathcal{P}$ could be significantly reduced during enumeration. 

For a fixed $\mathcal{P}$, we need to enumerate all possible $g_0$ and $g_1$. 
As $|x-g_0(x)|\leq T$ and $|x-g_1(x)|\leq T$, the time complexity to enumerate $g_0$ and $g_1$ is $O(\binom{2T+1}{2}^{|\mathcal{P}|})$. 
This requires us to choose small $|\mathcal{P}|$ and/or $T$, since the time complexity has the exponential form. 
Note that, for some $p\in \mathcal{P}$, there has no difference between $\{g_0(p) = x, g_1(p) = y\}$ and $\{g_0(p) = y, g_1(p) = x\}$ since the original message is always encrypted before embedding. That is why we use $\binom{2T+1}{2}^{|\mathcal{P}|}$ here, rather than $((2T+1)\cdot2T)^{|\mathcal{P}|}$.

From an empirical (or say heuristic) point of view, one can set $\mathcal{G}_0 = \mathcal{P}$ (i.e., $g_0(x) = x$), which has been utilized in traditional HS strategy. Thus, the time complexity to enumerate $g_0$ and $g_1$ is reduced as $O((2T+1)^{|\mathcal{P}|})$, which is significantly lower than the original one, yet still high. Actually, if we set $\mathcal{G}_0 = \mathcal{P}$, our task is to find optimal $g_1$ and $f$, which can be merged into the above optimization model. More general, once $g_0$ and $\mathcal{P}$ are fixed, we can find the optimal $g_1$ and $f$ out by calling the introduced weighted bigraph matching approach. Specifically, we will update $\mathcal{V}_1 = \mathcal{B}\setminus\mathcal{P}$ and 
$\mathcal{V}_2 = \mathcal{A}\setminus(\mathcal{G}_0\cup\mathcal{G}_1)$ as $\mathcal{V}_1 = \mathcal{B}$ and 
$\mathcal{V}_2 = \mathcal{A}\setminus \mathcal{G}_0$, respectively. Thus, the number of vertices in the corresponding weighted bigraph is $|\mathcal{B}| + |\mathcal{A}\setminus \mathcal{G}_0|$. 
Suppose that $\mathcal{V}_1 = \{u_1, u_2, ..., u_{|\mathcal{V}_1|}\}$ and $\mathcal{V}_2 = \{v_1, v_2, ..., v_{|\mathcal{V}_2|}\}$,
For every possible index-pair $(i, j)$, if $u_i\in \mathcal{V}_1\setminus\mathcal{P}$ and $|u_i-v_j| \leq T$, we then add an edge between $u_i$ and $v_j$, and the weight is determined as $C_{v_j-u_i}(u_i)$ according to Eq. (12).
Otherwise, if $u_i\in \mathcal{P} \subset \mathcal{V}_1$ and $|u_i-v_j| \leq T$, we then add an edge between $u_i$ and $v_j$, and the weight is determined as:
\begin{equation}
\begin{split}
C_{v_j-u_i}(u_i) &= \sum_{{e_l}^{(t)}=u_i}\delta({b_k}, 1)\cdot(v_j-u_i+{c_l}^{(t)}-{o_l}^{(t)})^2\\
&\approx \frac{1}{2}\cdot\sum_{{e_l}^{(t)}=u_i}(v_j-u_i+{c_l}^{(t)}-{o_l}^{(t)})^2,
\end{split}
\end{equation}
where $b_k=\{0,1\}$ means the \emph{k}-th bit to be embedded.
Note that, the difference between Eq. (12) and Eq. (16) is that, the PEH bins in Eq. (12) are shifted to ensure reversibility, while the PEH bins in Eq. (16) are shifted to hide message bits.

We take Fig. 2 for example. Let $\mathcal{P} = \mathcal{G}_0 = \{0, 1\}$. 
Fig. 4 (a) shows the new weighted bipartite graph, from which we can find new vertices and edges are added. Fig. 4 (b) shows an example of maximum matching. If it is optimal, then it means $g_1(0) = -1$ and $g_1(1) = 3$. And, the rest elements in $\mathcal{V}_1$ are also matched. Therefore, in applications, one can also enumerate all possible $\mathcal{P}$ and heuristically set $g_0$, e.g., $g_0(x)=x$, the optimal $g_1$ and $f$ can be then found by using the weighted bigraph matching method with a relatively low time complexity. Note that, for fixed $\mathcal{P}$ and $g_0$, one can find the optimal $g_1$ and $f$ out. Thereafter, he/she should further determine the global-optimal $\mathcal{P}$, $g_0$, $g_1$ and $f$ with Eq. (5, 6).

\subsection{Reversibility}
For reversibility, the data hider should self-embed the information of optimal $\mathcal{P}$, $g_0$, $g_1$ and $f$. As $|\mathcal{P}|$ is small, the space to store $\mathcal{P}$, $g_0$ and $g_1$ will be small. Note that, self-embedding $g_0$, $g_1$ and $f$ means to embed all required integer-pairs, e.g., $g_0(3) = 4$ tell us to self-embed (``3'',``4''). To self-embed $f: \mathcal{V}_1\mapsto \mathcal{V}_2$, one can first sort all elements in $\mathcal{V}_1$ in an increasing order, where the difference between any two adjacent elements in the ordered sequence is often small (e.g., ``-1'' in most cases) since the PEH is sharply distributed centered at zero-bin. This indicates that, we can use the run-length encoding (RLE) to lossless compress the differences. Meanwhile, the corresponding elements in $\mathcal{V}_2$ should be recorded as well. 
Let $\mathcal{V}_1$ be $\{u_1, u_2, ..., u_{|\mathcal{V}_1|}\}$, 
where $u_1 < u_2 < ... < u_{|\mathcal{V}_1|}$. we can compress $\{f(u_1)-u_1, f(u_2)-u_2, ..., f(u_{|\mathcal{V}_1|}) - u_{|\mathcal{V}_1|}\}$ by RLE or other efficient lossless algorithms since these differences are all bounded by a well-tuned $T$, e.g, $T = 1, 2$.

There are different methods to achieve self-embedding. For example, the data hider can choose a part of pixels (not in $\textbf{c}^{(t)}$) 
to store the above-mentioned auxiliary data.
The LSBs of these pixels will be kept unchanged throughout the specified-layer embedding such that one can successfully extract the hidden bits and recover the image content. The original LSBs will be considered as a part of the secret data.
\begin{figure}[!t]
\centering
\includegraphics[width=3in]{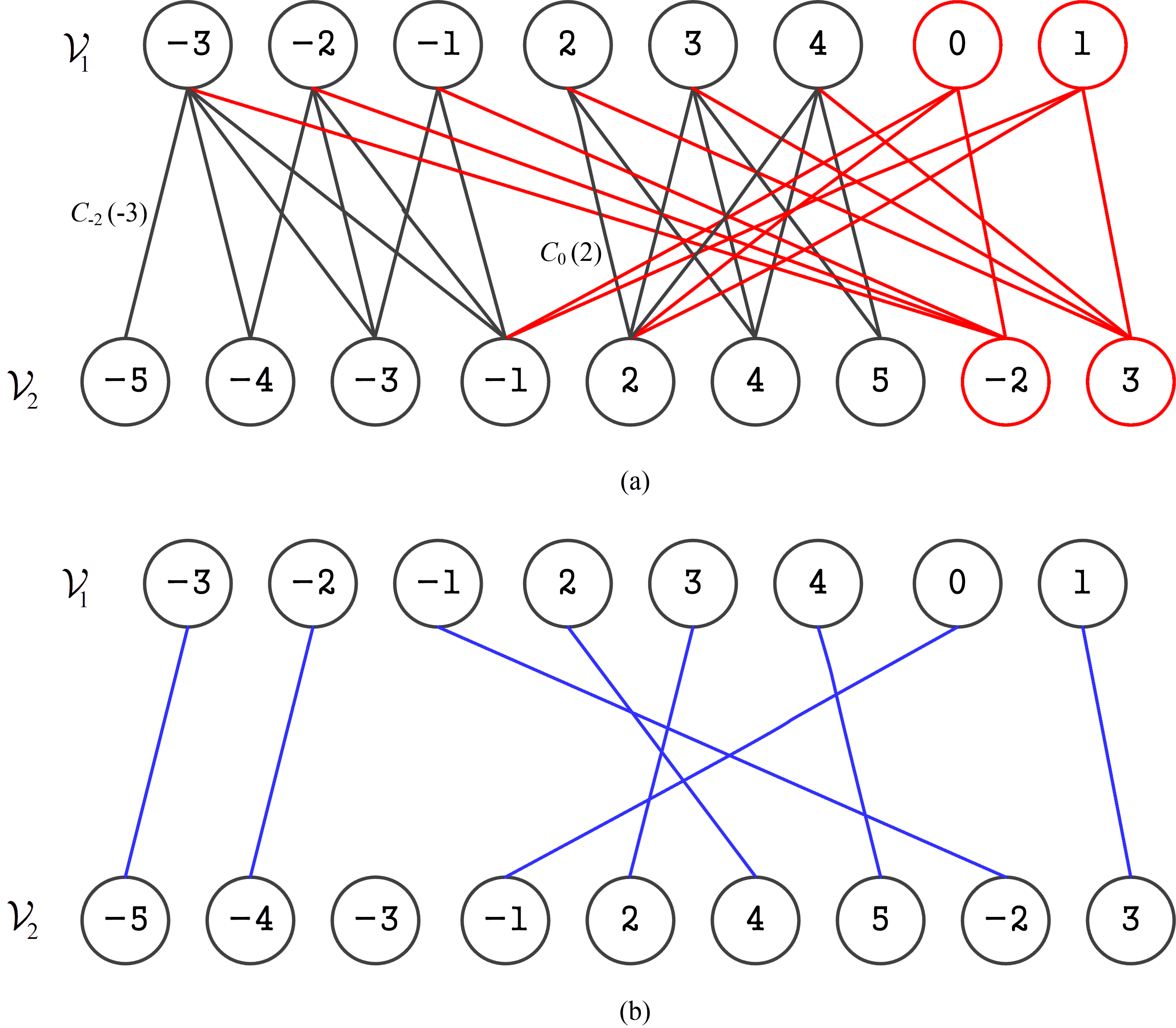}
\caption{An example to find both $g_1$ and $f$ for fixed $g_0$ and $\mathcal{P}$: (a) the weighted bipartite graph, (b) a maximum matching.}
\end{figure}
\section{Experimental Results and Analysis}
We incorporate the proposed optimization model into three state-of-the-art RDH algorithms, i.e., PC-HS \cite{tsai:pchs}, GF-HS Algorithm 1 \cite{li:GeneralHS} and DCSPF \cite{hzwu:DCSPF}, to evaluate the payload-distortion performance. In our experiments, for an RDH algorithm, we only optimize the data embedding operation with the proposed method, meaning that, the others such as pixel prediction, pixel selection and local-complexity function are all the same as the original ones. 
Since both PC-HS and DCSPF use PEH bin-pairs to carry the message bits, for fair comparison, we will set $|\mathcal{P}|=2$ for their optimized versions, denoted by ``PC-HS opt'' and ``DCSPF opt'', respectively. For simplicity, $|\mathcal{P}|$ is set to be 2 for ``GF-HS Algorithm 1 opt'' as well. Therefore, one can set any small $T > 0$ (e.g, $T = 1, 2$) since $2\cdot T\geq|\mathcal{P}|$ (to ensure that, a maximum matching can be always found).

In the PC-HS and GF-HS Algorithm 1, the data-hider has to use \emph{non-overlapped} pixel-blocks to carry the message bits. Here, the block-size is set to be $3\times 3$ for both methods, which is the same as described in the two methods. In the GF-HS Algorithm 1, before data embedding, the authors use a pixel selection parameter $s$ (that relies on the local-complexity function) to take advantage of smooth pixels as much as possible. In our simulation, when to use the proposed optimization method, there has no need to determine $s$ directly since we can sort the local-complexities in an increasing order such that smooth pixels can be utilized for data embedding, which is equivalent to using $s$. In the DCSPF, the data-hider needs to set two important parameters, namely the pixel-blocking rate and the number of selection-layers. As recommended in the method, we enumerate the pixel-blocking rate from 10\% to 90\% with a step of 10\%, and vary the number of selection-layers from 3 to 6 with a step of 3.

During data embedding, the \emph{multiple-pass embedding} strategy \cite{li:GeneralHS} is applied for both PC-HS and GF-HS Algorithm 1. Additionally, for a required payload, a given image may be embedded several times (namely called \emph{multi-layer embedding}). For multi-layer embedding, since it is free to set the payload size of each layer, in default, we will embed message bits into a specified layer as much as possible with a payload-step until it cannot carry additional bits, and a higher-layer embedding is then applied. Note that, this strategy may be not optimal.
\begin{figure*}[!t]
\centering
\includegraphics[width=6.5in]{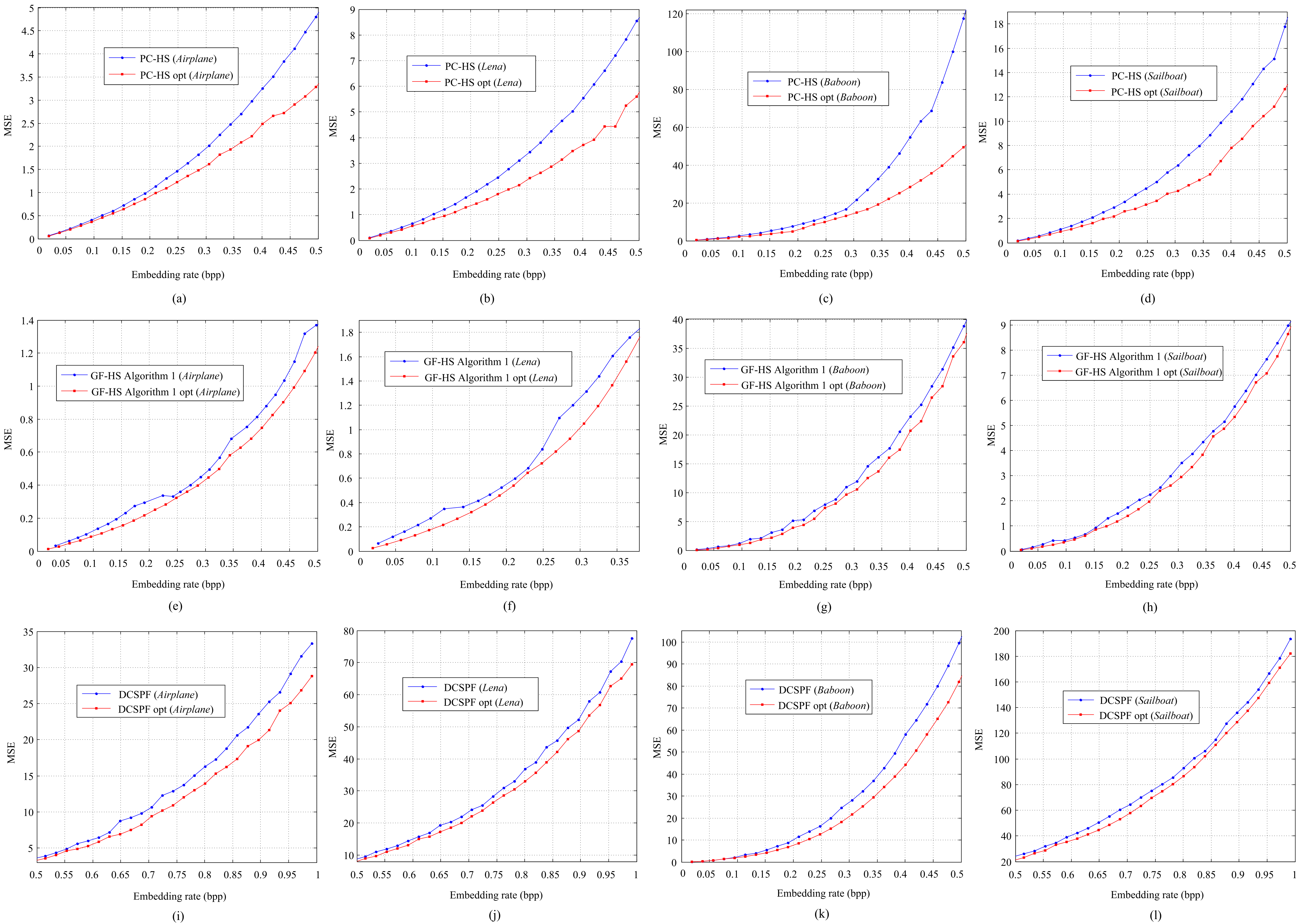}
\caption{The payload-distortion performance comparison for different RDH algorithms with/without the proposed optimization method.}
\end{figure*}

We here take four grayscale images \emph{Airplane}, \emph{Lena}, \emph{Baboon} and \emph{Sailboat} sized $512\times 512\times 8$ for experiments. The MSE defined in Eq. (2) is used as the distortion measure. Fig. 5 shows the payload-distortion performance for different RDH algorithms with/without the proposed optimization method. It can be seen from Fig. 5 that, our optimization method has the property to significantly reduce the introduced distortion due to data embedding, implying that, the proposed method could be promising for both practical use and the RDH design. It is noted that, for DCSPF, when the embedding rate is lower than 0.5 bpp, the performance improvement is not significant for \emph{Airplane}, \emph{Lena} and \emph{Sailboat} since the authors also use an efficient approximation algorithm to find near-optimal PEH bin-pairs. It indicates that, when $|\mathcal{P}|=2$, to a certain extent, the approximation algorithm proposed by Wu \emph{et al.} \cite{hzwu:DCSPF} can be used as an approximate solution of the proposed model.
\section{Conclusion and Discussion}
In this paper, we have proven that, the traditional HS operation corresponds to a maximum matching in the corresponding bigraph. To reduce the embedding distortion, based on $\textbf{x}^{(0)}$ and $\textbf{x}^{(t)}$, we model the HS operation as a minimum weighted matching problem, and use the MWMM technique to find the best HS strategy for RDH. We incorporate our optimization model into some related works and experimental results have shown that the optimization method can improve the payload-distortion performance. For the proposed optimization model, in applications, the time complexity to enumerate all usable $\mathcal{P}$ would be still very high for a large $|\mathcal{P}|$. In the future, we expect to study heuristic algorithms to find near-optimal $\mathcal{P}$.



\bibliographystyle{IEEEtran}
\bibliography{IEEEexample}
%

%
%
%

\end{document}